\documentclass[english,showpacs,preprintnumbers,prbr]{revtex4-1}
\usepackage[latin9]{inputenc}
\usepackage{amsmath}
\usepackage{amssymb}
\usepackage{graphicx}
\usepackage{esint}
\usepackage{natbib}
\usepackage{hyperref}
\hypersetup{colorlinks=true,linkcolor=blue,filecolor=magenta,urlcolor=blue,citecolor=blue,}

\makeatletter

\@ifundefined{textcolor}{}{%
 \definecolor{BLACK}{gray}{0}
 \definecolor{WHITE}{gray}{1}
 \definecolor{RED}{rgb}{1,0,0}
 \definecolor{GREEN}{rgb}{0,1,0}
 \definecolor{BLUE}{rgb}{0,0,1}
 \definecolor{CYAN}{cmyk}{1,0,0,0}
 \definecolor{MAGENTA}{cmyk}{0,1,0,0}
 \definecolor{YELLOW}{cmyk}{0,0,1,0}
}

\usepackage{color}
\usepackage{ulem}

\usepackage{aecompl}

\usepackage{epsfig}\usepackage{dcolumn}\usepackage{bm}

\usepackage{babel}

\makeatother

\usepackage{babel}
\begin{document}

\title{Tuning of heat and charge transport by Majorana fermions}

\author{L. S. Ricco$^{1}$, F. A. Dessotti$^{1}$, I. A. Shelykh$^{2,3}$, M. S. Figueira$^{4}$, and A. C. Seridonio$^{1,5}$}

\email[correspondent author: ]{seridonio@dfq.feis.unesp.br}

\affiliation{$^{1}$Departamento de F\'{i}sica e Qu\'{i}mica, Unesp - Univ Estadual Paulista, 15385-000, Ilha Solteira, SP, Brazil\\
$^{2}$Science Institute, University of Iceland, Dunhagi-3, IS-107, Reykjavik, Iceland\\
$^{3}$ITMO University, St. Petersburg 197101, Russia\\
$^{4}$Instituto de F\'{i}sica, Universidade Federal Fluminense, 24210-340, Niter\'{o}i, RJ, Brazil\\
$^{5}$IGCE, Unesp - Univ Estadual Paulista, Departamento de F\'{i}sica, 13506-900, Rio Claro, SP, Brazil}

\begin{abstract}
We investigate theoretically thermal and electrical conductances for the system consisting of a quantum dot (QD) connected both to a pair of Majorana fermions residing the edges of a Kitaev wire and two metallic leads. We demonstrate that both quantities reveal pronounced resonances, whose positions can be controlled by tuning of an asymmetry of the couplings of the QD and a pair of MFs. Similar behavior is revealed for the thermopower, Wiedemann-Franz law and dimensionless thermoelectric figure of merit. The considered geometry can thus be used as a tuner of heat and charge transport assisted by MFs.
\end{abstract}
\maketitle

\section*{Introduction} \label{sec1}

\hspace{0.3cm} Majorana fermions (MFs) are particles that are equivalent to their antiparticles. The corresponding concept was first proposed in the domain of high-energy physics, but later on existence of the elementary excitations of this type was predicted for certain condensed matter systems. Particularly, MFs emerge as quasiparticle excitations characterized by zero-energy modes\cite{Alicea,Franz} appearing at the edges of the 1D Kitaev wire\cite{Kitaev,Kitaev1,Kitaev2,Kitaev3,Kitaev4}. Kitaev model is used to describe the emerging phenomena of $p$-wave and spinless topological superconductivity.

Kitaev topological phase can be experimentally achieved in the geometry consisting of a semiconducting nanowire with spin-orbit interaction put in contact with $s$-wave superconducting material and placed in external magnetic field\cite{wire1,wire2016}. Other condensed matter systems were also proposed as candidates for the observation of MFs. They include ferromagnetic chains placed on top of superconductors with spin-orbit interaction\cite{wire2,Jelena2}, fractional quantum Hall state with filling factor $\nu=5/2$ \cite{QH}, three-dimensional topological insulators\cite{TI} and superconducting vortices\cite{V1,V2,V3}.

\begin{figure}[!]
\begin{center}
\includegraphics[width=0.55\textwidth,height=0.3\textheight]{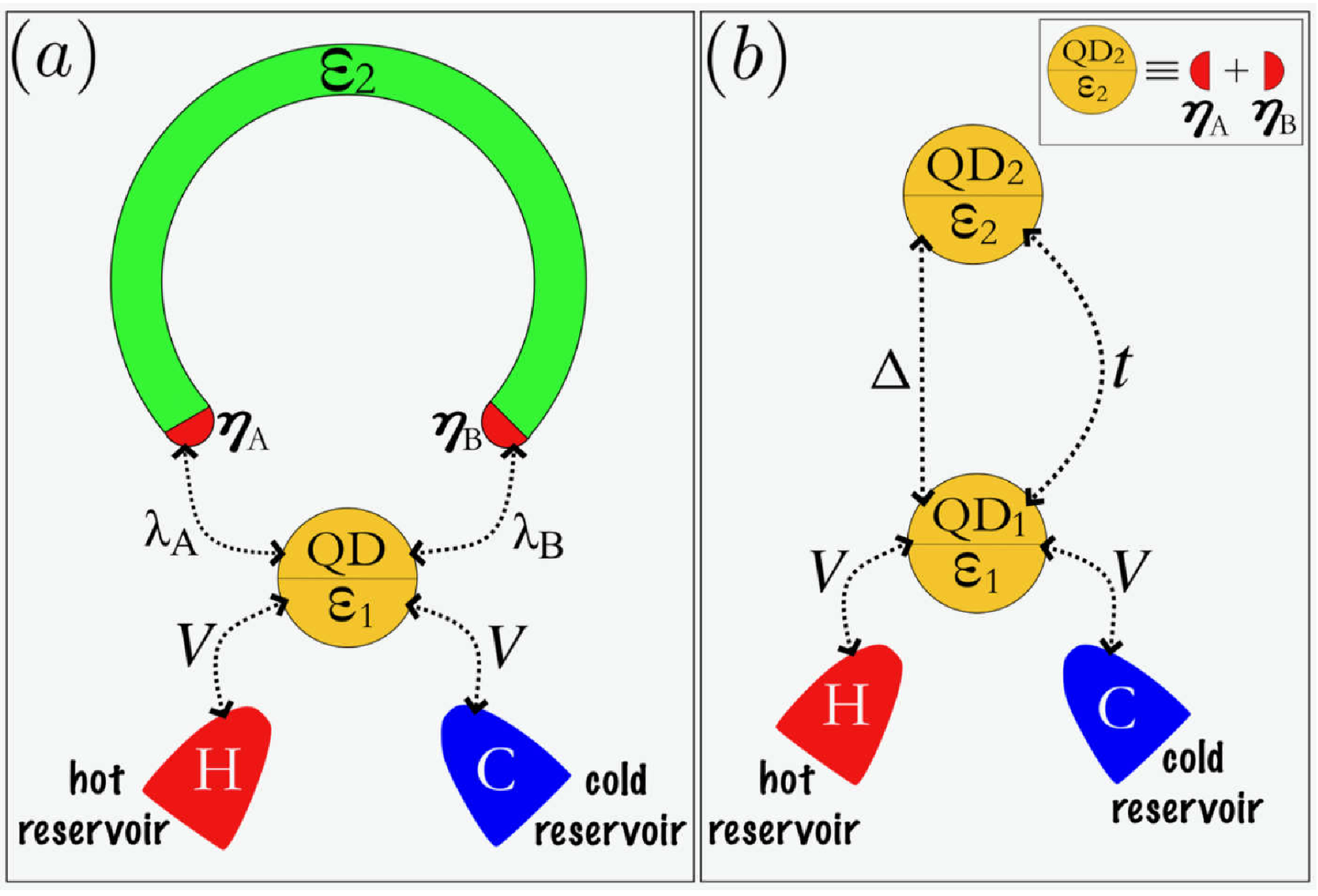} \caption{\label{fig:1}(a) The sketch of the geometry we consider. Topological U-shaped Kitaev wire with a pair of MFs $\eta_{A}$ and $\eta_{B}$ is placed in contact with a QD, which is connected as well to two metallic reservoirs. The coupling of the QD to the MFs is asymmetric and is characterized by tunneling matrix elements $\lambda_{A}$ and $\lambda_{B}$, while coupling to the metallic leads is symmetric and is characterized by the tunneling matrix element $V$. $\varepsilon_{2}$ denotes the coupling between two MF states. (b) Equivalent auxiliary setup (Kitaev dimer) resulting from the mapping of the original system onto the system with nonlocal fermion residing in $\text{QD}_{2}$. $t$ is tunneling matrix element between the QDs 1 and 2, $\Delta$ is the binding energy of the Cooper pair delocalized between them.}
\end{center}
\end{figure}

MFs residing at the opposite edges of a Kitaev wire are elements of a robust nonlocal \textit{qubit} which appears to be immune to the environment decoherence. This attracted the interest of the researchers working in the domain of quantum information and transport, as systems with MFs \cite{Flens1,Flens2,Ring2017} can be in principle used as building blocks for the next generation of nanodevices, \cite{Dessotti,Ricco} including current switches \cite{Dessotti} and quantum memory elements\cite{Ricco}. At the same time, similar systems were proposed as thermoelectric nanodevices \cite{Orellana,RLopez,Lei,Valentini}.

In this work, following the proposals of thermoelectric detection of MF states \cite{Orellana,RLopez,Lei,Valentini}, we explore theoretically zero-bias thermal and electrical transport through one particular geometry consisting of an individual QD coupled both to a pair of MFs and metallic leads as shown in the Fig.\ref{fig:1}(a). The MFs reside at the edges of a topological U-shaped Kitaev wire, similar to the case of Ref.[19]. The QD coupling to the MFs is considered to be asymmetric, while coupling to the metallic leads is symmetric, and MFs are supposed to overlap with each other. The results of our calculations clearly show that thermoelectric conductance, thermopower, Wiedemann-Franz law\cite{Ventra} and dimensionless thermoelectric figure of merit (ZT) as function of the QD  electron energy demonstrate resonant behavior. Moreover, the position of the resonance can be tuned by changing the coupling amplitudes between the QD and the MFs, which allows the system to operate as a tuner of heat and charge assisted by MFs.

\section*{Model}\label{sec2}

\hspace{0.3cm} For theoretical treatment of the setup depicted in the Fig.\,\ref{fig:1}(a), we use the Hamiltonian proposed by Liu and Baranger\cite{Baranger}:
\begin{eqnarray}
\mathcal{H} & = & \underset{\alpha k}{\sum}\varepsilon_{k}c_{\alpha k}^{\dagger}c_{\alpha k}+\varepsilon_{1}d_{1}^{\dagger}d_{1}+V\underset{\alpha k}{\sum}(c_{\alpha k}^{\dagger}d_{1}+\text{{H.c.}}) + {\lambda_{A}}(d_{1}-d_{1}^{\dagger})\eta_{A}+{\lambda_{B}}(d_{1}+d_{1}^{\dagger})\eta_{B} + i\varepsilon_{2}\eta_{A}\eta_{B}, \label{eq:TIAM}
\end{eqnarray}
where the electrons in the leads $\alpha=H,C$ (for hot and cold reservoirs,
respectively) are described by the operators $c_{\alpha k}^{\dagger}$
($c_{\alpha k}$) for the creation (annihilation) of an electron in
a quantum state labeled by the wave number $k$ and energy $\varepsilon_{k}$. For the QD $d_{1}^{\dagger}$ ($d_{1}$) creates (annihilates) an electron
in the state with the energy $\varepsilon_{1}.$ The energies of both electrons in the leads and QD are counted from the chemical potential $\mu$ (we consider only the limit of small source-drain bias, thus assuming that chemical potential is the same everywhere). $V$ stands for the hybridization between the QD and the leads. The asymmetric coupling between the QD and MFs at the edges of the topological U-shaped Kitaev wire is described by the complex tunneling amplitudes $\lambda_{A}$ and $\lambda_{B}$. Introduction of an asymmetry in the couplings can account for the presence of the magnetic flux which can be introduced via Peierls phase shift \cite{Baranger}. $\varepsilon_2$ stands for the overlap between the MFs.

Without the loss of generality, we can put: $\lambda_{A}=\frac{(t+\Delta)}{\sqrt{2}}$ and $\lambda_{B}=i\frac{(\Delta-t)}{\sqrt{2}},$ respectively for the left $(\eta_{A}=\eta_{A}^\dagger)$ and right $(\eta_{B}=\eta_{B}^\dagger)$ MFs, and introduce an auxiliary nonlocal fermion $d_{2}=\frac{1}{\sqrt{2}}(\eta_{A}+i\eta_{B})$ \cite{Dessotti,Ricco}. {The expressions for $\lambda_{A}=|\lambda_{A}|e^{i\phi_{A}}$ and $\lambda_{B}=|\lambda_{B}|e^{i\phi_{B}}$ constitute a convenient gauge for our problem. We put $\phi_{A}=0$ and $\phi_{B}=(n+\frac{1}{2})\pi$ with integer $n=0,1,2,\ldots$ corresponding to the total flux through the ring of Fig.\,\ref{fig:1}. This parameter is experimentally tunable by changing the external magnetic field. This fact gives certain advantages to our proposal with respect to the previous works with asymmetric couplings between a single QD and a pair of MFs at the ends of a topological Kitaev wire\cite{DJClarke,PRecher,RAguado,RAguadoExp}. According to Ref.[32] the parameter $\varepsilon_2$ describing the overlap between the MFs depends on magnetic field in an oscillatory manner, the amplitudes $|\lambda_{A}|=\frac{t+\Delta}{\sqrt{2}}$ and $|\lambda_{B}|=\frac{|\Delta-t|}{\sqrt{2}}$ demonstrate the same behavior (see Sec.III-A of Ref.[30]) and thus external magnetic field affects not only the relative phase between $\lambda_A$ and $\lambda_B$ but their absolute values as well. To fulfill the condition $|\lambda_{B}|<|\lambda_{A}|$ one should place the QD closer the MF $\eta_{A}$ than to the MF $\eta_{B}$.}

We map the original Hamiltonian into one where the electronic states $d_{1}$ and $d_{2}$ are connected via normal tunneling $t$ and bounded as delocalized Cooper pair, with binding energy $\Delta$:

\begin{eqnarray}
\mathcal{H} & = & \underset{\alpha k}{\sum}\varepsilon_{k}c_{\alpha k}^{\dagger}c_{\alpha k}+ V\underset{\alpha k}{\sum}(c_{\alpha k}^{\dagger}d_{1}+\text{{H.c.}}) + \varepsilon_{1}d_{1}^{\dagger}d_{1} +  \varepsilon_{2}d_{2}^{\dagger}d_{2}+(td_{1}d_{2}^{\dagger}+\Delta d_{2}^{\dagger}d_{1}^{\dagger}+\text{{H.c.}}) -\frac{\varepsilon_2}{2}. \label{NewHamiltonian}
\end{eqnarray}

This expression represents a shortened version of the microscopic model for the Kitaev wire corresponding to the Kitaev dimer (see Fig.\ref{fig:1}(b)). As it was shown in the Refs.[33] and [34] this model allows clear distinguishing between topologically trivial and Majorana-induced zero-bias peak in the conductance.

In what follows, we use the Landauer-B\"{u}ttiker formula for the zero-bias
thermoelectric quantities $\mathcal{L}_{n}$\cite{RLopez,Orellana}:
\begin{equation}
\mathcal{L}_{n}=\frac{1}{h}\int d\varepsilon\left(-\frac{\partial f_{F}}{\partial\varepsilon}\right)\varepsilon^{n}\mathcal{T},\label{eq:G}
\end{equation}
where $h$ is Planck's constant, $\Gamma=2\pi V^{2}\sum_{k}\delta(\varepsilon-\varepsilon_{k})$
is Anderson broadening\cite{Anderson} and $f_{F}$ stands for Fermi-Dirac distribution. The quantity
\begin{equation}
\mathcal{T}=-\Gamma\text{{Im}}(\tilde{\mathcal{G}}_{d_{1}d_{1}}) \label{eq:Transmittance}
\end{equation}
is electronic transmittance through the QD, with $\tilde{\mathcal{G}}_{d_{1}d_{1}}$ being
retarded Green's function for the QD in the energy domain $\varepsilon,$
obtained from the Fourier transform $\tilde{\mathcal{G}}_{\mathcal{\mathcal{A}B}}=\int d\tau\mathcal{G}_{\mathcal{\mathcal{A}B}}e^{\frac{i}{\hbar}(\varepsilon+i0^{+})\tau},$
where
\begin{equation}
\mathcal{G}_{\mathcal{AB}}=-\frac{i}{\hbar}\theta(\tau){\tt Tr}\{\varrho[\mathcal{A}(\tau),\mathcal{B}^{\dagger}(0)]_{+}\} \label{eq:GF}
\end{equation}
corresponds to the Green's function in time domain $\tau,$ expressed
in terms of the Heaviside function $\theta\left(\tau\right)$ and thermal density matrix $\varrho$ for Eq.\,(\ref{eq:TIAM}).

Experimentally measurable thermoelectric coefficients can be expressed via $\mathcal{L}_{0},\mathcal{L}_{1}$ and $\mathcal{L}_{2}$ as:
\begin{equation}
G=e^{2}\mathcal{L}_{0},\label{eq:Gelec}
\end{equation}
\begin{equation}
K=\frac{1}{T}(\mathcal{L}_{2}-\frac{\mathcal{L}_{1}^{2}}{\mathcal{L}_{0}})\label{eq:K}
\end{equation}
and
\begin{equation}
S=-(\frac{1}{eT})\frac{\mathcal{L}_{1}}{\mathcal{L}_{0}} \label{eq:S}
\end{equation}
for the electrical and thermal conductances and thermopower, respectively (T denotes a temperature of the system).

We also investigate the violation of Wiedemann-Franz law, given by
\begin{equation}
 WF=\frac{1}{T}(\frac{K}{G}),\label{eq:WF}
\end{equation}
in units of Lorenz number $L_{0}=(\pi^{2}/3)(k_{B}/e)^{2}$ and corresponding behavior of the dimensionless figure of merit \cite{RLopez,Orellana}
\begin{equation}
ZT=\frac{S^{2}GT}{K}.\label{eq:ZT}
\end{equation}

For Eq.\,(\ref{eq:Transmittance}), we use equation-of-motion (EOM) method\cite{book} summarized as follows:
\begin{equation}
(\varepsilon+i0^{+})\tilde{\mathcal{G}}_{\mathcal{AB}}=[\mathcal{A},\mathcal{B^{\dagger}}]_{+}+{\tilde{\mathcal{G}}_{\left[\mathcal{A},\mathcal{\mathcal{H}}\right]\mathcal{B}}}, \label{eq:EOM}
\end{equation}
with $\mathcal{A=B}=d_{1}$.

As our Hamiltonian given by Eqs.~(\ref{eq:TIAM}) and (\ref{NewHamiltonian}) is quadratic, the set of the EOM for the single particle Green's functions can be closed without any truncation procedure \cite{Ricco2}. We find the following four coupled linear algebraic equations:
\begin{align}
(\varepsilon-\varepsilon_{1}-\Sigma)\tilde{\mathcal{G}}_{d_{1}d_{1}} & =1-t\mathcal{\tilde{G}}_{d_{2}d_{1}}-\Delta\mathcal{\tilde{G}}_{d_{2}^{\dagger}d_{1}},\label{eq:GF11}
\end{align}
where $\Sigma=-i\Gamma$ is the self-energy of the coupling with the metallic leads
\begin{eqnarray}
\mathcal{\tilde{G}}_{d_{2}d_{1}} & = & +\frac{\Delta\mathcal{\tilde{G}}_{d_{1}^{\dagger}d_{1}}}{(\varepsilon-\varepsilon_{2}+i0^{+})}-\frac{t\mathcal{\tilde{G}}_{d_{1}d_{1}}}{(\varepsilon-\varepsilon_{2}+i0^{+})},\label{eq:GF21}\\
\mathcal{\tilde{G}}_{d_{2}^{\dagger}d_{1}} & = & -\frac{\Delta\mathcal{\tilde{G}}_{d_{1}d_{1}}}{(\varepsilon+\varepsilon_{2}+i0^{+})}+\frac{t\mathcal{\tilde{G}}_{d_{1}^{\dagger}d_{1}}}{(\varepsilon+\varepsilon_{2}+i0^{+})}\label{eq:GF21plus}
\end{eqnarray}
and
\begin{equation}
\mathcal{\tilde{G}}_{d_{1}^{\dagger}d_{1}}=-2t\Delta\tilde{K}\mathcal{\tilde{G}}_{d_{1}d_{1}},\label{eq:Gd1dagd1}
\end{equation}
with
\begin{equation}
\tilde{K}=\frac{K_{\text{{MFs}}}}{\varepsilon+\varepsilon_{1}-\Sigma-K_{-}},\label{eq:Ktilde}
\end{equation}
\begin{equation}
K_{\text{{MFs}}}=\frac{(\varepsilon+i0^{+})}{[\varepsilon^{2}-\varepsilon_{2}^{2}+2i\varepsilon0^{+}-(0^{+})^{2}]}\label{eq:KMFs}
\end{equation}
and
\begin{equation}
K_{\pm}=\frac{(\varepsilon+i0^{+})(t^{2}+\Delta^{2})\pm\varepsilon_{2}(t^{2}-\Delta^{2})}{[\varepsilon^{2}-\varepsilon_{2}^{2}+2i\varepsilon0^{+}-(0^{+})^{2}]}.\label{eq:Kpm}
\end{equation}

This gives the Green's function of the QD:
\begin{align}
\tilde{\mathcal{G}}_{d_{1}d_{1}} & =\frac{1}{\varepsilon-\varepsilon_{1}-\Sigma-\Sigma_{\text{{MFs}}}},\label{eq:d1d1}
\end{align}
where the part of self-energy
\begin{equation}
\Sigma_{\text{{MFs}}}=K_{+}+(2t\Delta)^{2}\tilde{K}K_{\text{{MFs}}}\label{eq:SIGMA_MFs}
\end{equation} describes the hybridization between MFs and QD.

Importantly, for the low temperatures regime, the substitution of  Eq.\,(\ref{eq:d1d1}) into Eq.\,(\ref{eq:G}) and its decomposition into Sommerfeld series \cite{RLopez,Ventra} allows to get analytical expressions for thermoelectric coefficients:
\begin{equation}
\frac{G}{G_{0}}=\frac{K}{G_{0}L_{0}T}\approx\left.\mathcal{T}\right|_{\varepsilon=0}\label{eq:GK},
\end{equation}
\begin{equation}
S\approx eL_{0}T\left.\frac{1}{\mathcal{T}}\frac{d\mathcal{T}}{d\varepsilon}\right|_{\varepsilon=0},\label{eq:Saprox}
\end{equation}
where
\begin{equation}
\mathcal{T}=\frac{\tilde{\Gamma}^{2}}{[\varepsilon-\varepsilon_{1}-K_{+}-\frac{(2t\Delta K_{\text{MFs}})^{2}(\varepsilon+\varepsilon_{1}-K_{-})}{(\varepsilon+\varepsilon_{1}-K_{-})^{2}+\Gamma^{2}}]{}^{2}+\tilde{\Gamma}^{2}},\label{eq:Trans}
\end{equation}
with
\begin{equation}
\tilde{\Gamma}=[1+\frac{(2t\Delta K_{\text{MFs}})^{2}}{(\varepsilon+\varepsilon_{1}-K_{-})^{2}+\Gamma^{2}}]\Gamma. \label{eq:Broad}
\end{equation}

Comparison of the Eqs.~(\ref{eq:GK}) and (\ref{eq:Saprox}) allows us to conclude that the peak values of the electric conductance are reached when $S=0$ for which $d\mathcal{T}/d\varepsilon=0$ which happens when
\begin{equation}
\varepsilon_{1}=\frac{(t^{2}-\Delta^{2})}{\varepsilon_{2}}.\label{eq:Positions}
\end{equation}

As we will see below, fulfillment of this condition corresponds to the presence of an electron-hole symmetry in the system. Note that as $\varepsilon_2$ enters in the denominator of the Eq.~(\ref{eq:Positions}), even slight differences between $t$ and $\Delta$ will be enough to change drastically the position of the resonance if hybridization between the MFs is small.

\section*{Results and Discussion} \label{sec3}

\hspace{0.3cm} In our further calculations, we scale the energy in units of the Anderson broadening
$\Gamma=2\pi V^{2}\sum_{k}\delta(\varepsilon-\varepsilon_{k})$\cite{Anderson} and take the temperature of the system $k_{B}T=10^{-4}\Gamma$. The Anderson broadening $\Gamma$ defines the coupling between the QD and the metallic leads, which is assumed to be symmetrical for a sake of simplicity.

We start our analysis from the case when only a single MF ($\eta_{A}$) is coupled to the QD. In terms of the amplitudes $t,\Delta$ this corresponds to $t=\Delta$. To be specific, we fix $t=\Delta=4\Gamma$. Looking at Eq.~(\ref{NewHamiltonian}), we see that the terms $d_{1}d_{2}^{\dagger}+\text{{H.c.}}$ and $d_{2}^{\dagger}d_{1}^{\dagger}+\text{{H.c.}}$ enter into Hamiltonian with equal weights, and thus we are in the superconducting (SC)-metallic boundary phase.

\begin{figure}[!]
\begin{center}
\includegraphics[width=0.55\textwidth,height=0.3\textheight]{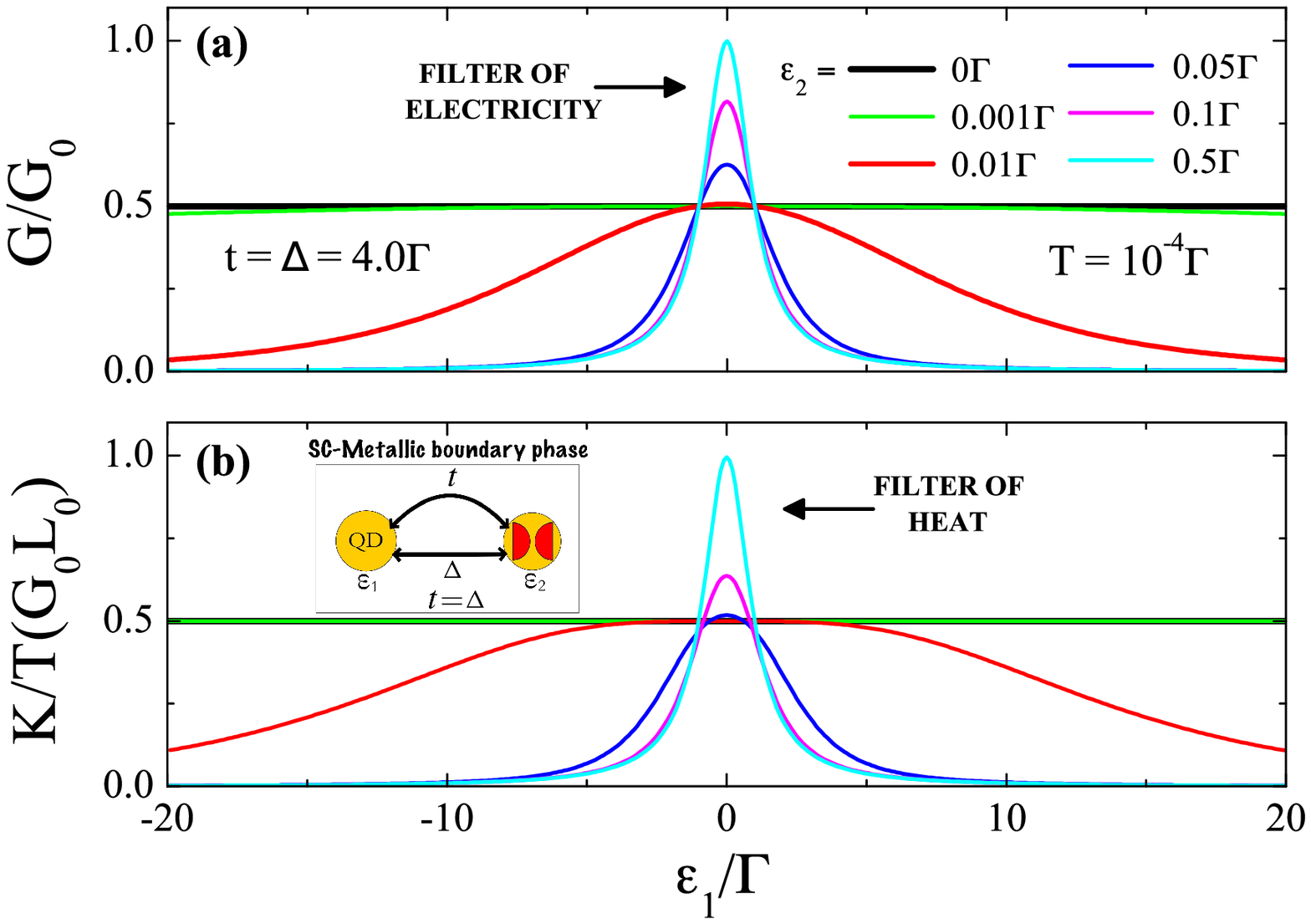}
\caption{\label{fig:2} Electrical and thermal conductances of the system corresponding to SC-metallic boundary phase, $t=\Delta=4\Gamma$: (a) Electrical conductance
as function of the QD energy level $\varepsilon_{1}$ for several $\varepsilon_{2}$
values of the couplings between MFs. (b) Corresponding thermal conductance. For both cases the resonance at the Fermi energy $\varepsilon_{1}=0$ occurs if $\varepsilon_{2}\protect\neq0$. For $\varepsilon_{2}\protect=0$ the conductance plateau is observed (see main text for the corresponding discussion). The inset shows the equivalent circuit with an auxiliary fermion $d_{2}$ constructed from MFs $\eta_{A}$ and $\eta_{B}$ (red half-circles).}
\end{center}
\end{figure}

\begin{figure}[!]
\begin{center}
\includegraphics[width=0.55\textwidth,height=0.3\textheight]{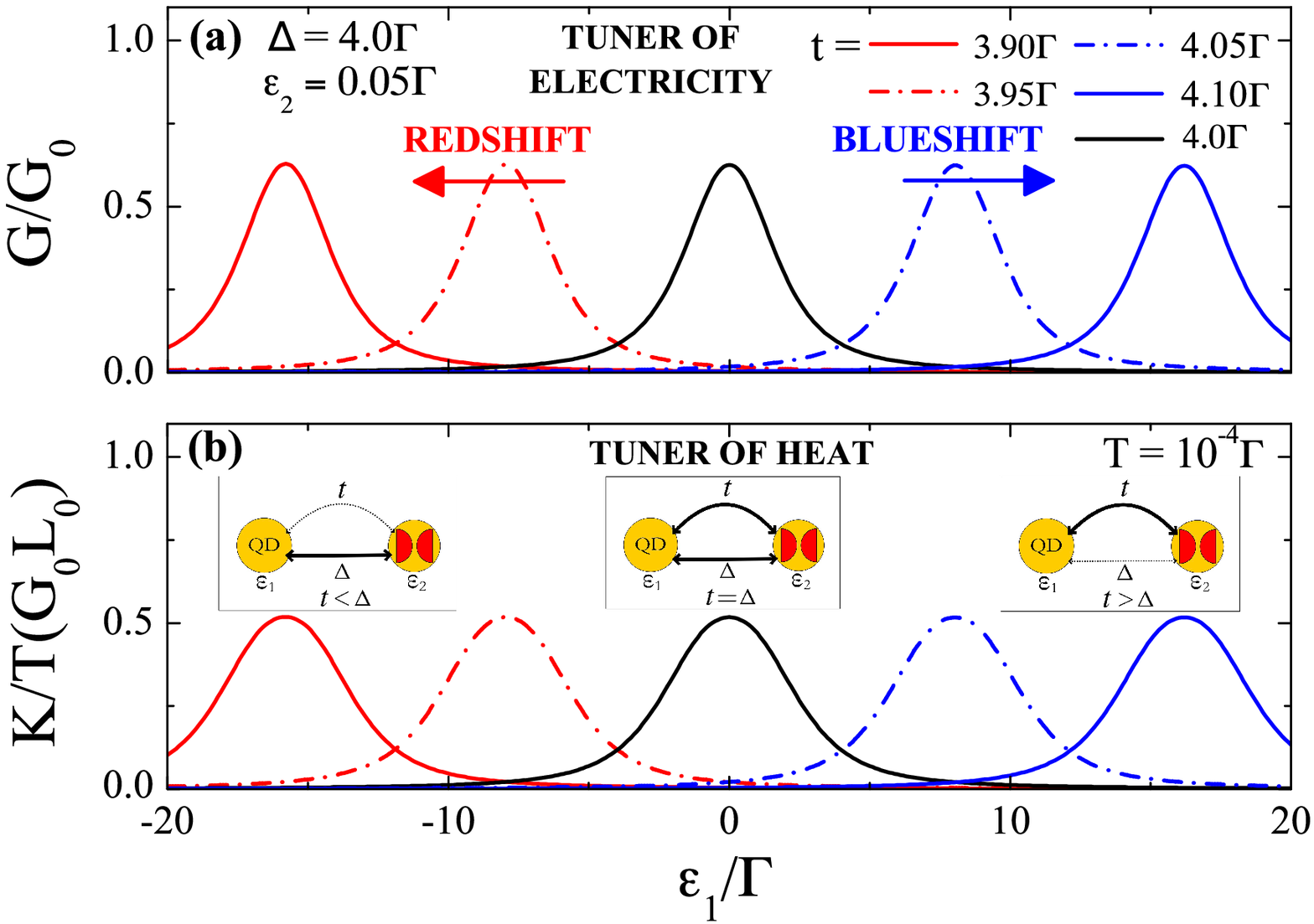}
\caption{\label{fig:3} Electrical and thermal conductances as functions of the QD energy level outside SC-metallic boundary phase. Slight deviations from the condition $t=\Delta,$ result in the shift of the resonance peak for the electrical (panel (a)) and thermal (panel (b)) conductances. The corresponding resonances are blueshifted for $t>\Delta$ and redshifted for $t<\Delta$ as compared to the case of the SC-metallic boundary phase. Insets show the equivalent circuit with auxiliary fermion $d_{2}$ constructed from MFs $\eta_{A}$ and $\eta_{B}$ (red half-circles).}
\end{center}
\end{figure}

Fig.\ref{fig:2}(a) shows the electrical conductance $G=e^{2}\mathcal{L}_{0}$ scaled in units of the conductance quantum $G_{0}=e^{2}/h$ as a function of the QD energy level  $\varepsilon_{1}$, for several coupling amplitudes $\varepsilon_2$ between the MFs. Note that, if MFs are completely isolated from each other ($\varepsilon_{2}=0$), the conductance reveals a plateau with $G=G_{0}/2$ whatever the value of $\varepsilon_{1}$ (black line), and similar trend is observed in the thermal conductance shown in the Fig.\,\ref{fig:2}(b). The effect is due to the leaking of the Majorana fermion state into the QD\cite{Vernek}. The MF zero-mode becomes pinned at the Fermi level of the metallic leads, but within the QD electronic-structure. With increase of the coupling between the wire and the QD, the MF state of the Kitaev wire leaks into the QD. As a result, a peak at the Fermi energy emerges in the QD density of states (DOS), while in the DOS corresponding to the edge of the wire the corresponding peak becomes gradually suppressed. Consequently, the QD effectively becomes the new edge of the Kitaev wire. This scenario was reported experimentally in the Ref.[9].

To get resonant response of the thermoelectric conductances one should consider the case $\varepsilon_{2}\neq0$, corresponding to the splitting of the MF zero-bias peak.  The resonant behavior of $G$ and $K$ can be understood as arising from the presence of an auxiliary fermion $d_{2},$ in the Hamiltonian [Eq.~(\ref{NewHamiltonian})], whose energy $\varepsilon_{2}$ is now detuned from the Fermi level (see inset of Fig.\,\ref{fig:2}(b)). In this case, the regular fermion state instead of the corresponding half-fermion provided by MF $\eta_{A}$ gives the main contribution to the charge and heat current. In this scenario, filtering of the electricity and heat emerges: the maximal transmission occurs at $\varepsilon_{1}=0$. { Our Figs.\ref{fig:2}(a) and (b) recover the findings of Fig.5(a) in Ref.[23]. Our work, however, have an important novel dimension: we demonstrate that even small deviations of the system from the SC-metallic boundary phase which can be achieved by the control of the asymmetry of the couplings allows realization of the efficient tuners of electricity and heat.} This effect is shown in the Figs.\,\ref{fig:3}(a) and (b). As one can see, even small detuning of the coefficient $t$ from the value $t=\Delta$ leads to substantial blueshift (for the case $t>\Delta$) or redshift (for the case $t<\Delta$) of the conductance resonances. Such sensitivity is a direct consequence of the Eq.~(\ref{eq:Positions}) defining the position of the resonances.

\begin{figure}[!]
\begin{center}
\includegraphics[width=0.55\textwidth,height=0.3\textheight]{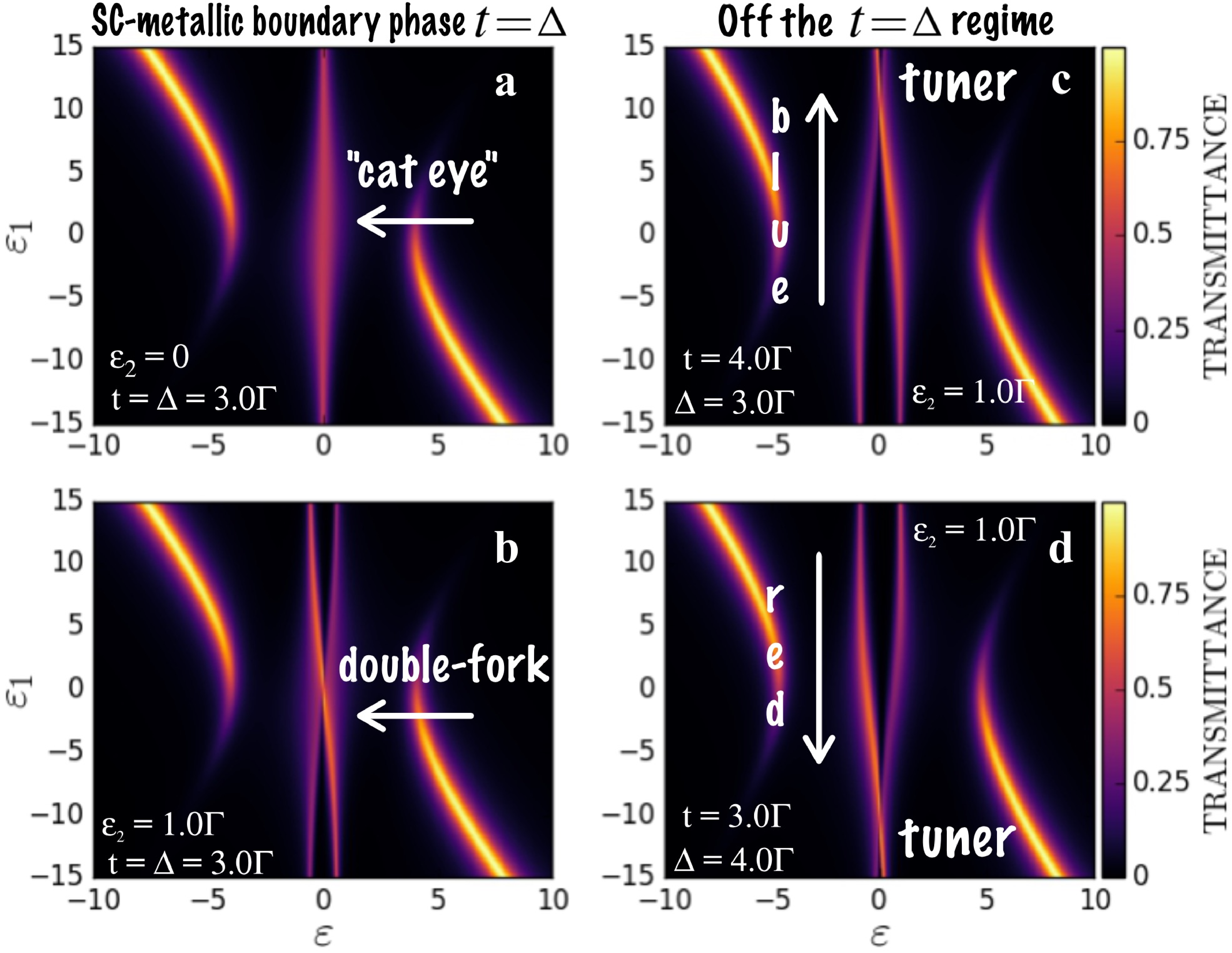}
\caption{\label{fig:4}Transmittance $\mathcal{T}$ spanned by the axes of $\varepsilon_{1}$ and $\varepsilon$. Panels (a) and (b) show the regime corresponding to SC-metallic boundary phase with $t=\Delta,$ for $\varepsilon_{2}=0$ and finite $\varepsilon_{2}$, respectively. Panel (a) reveals characteristic ``cat eye"-shaped central structure at the Fermi level responsible for the onset of the conductance plateau. Panel (b) exhibits a double-fork structure responsible for the resonant character of the conductance for $\varepsilon_2\neq0$. Introduction of the asymmetry of the QD to MFs coupling leads to the vertical shift of the double-fork feature resulting in the blueshift (panel (c)) or redshift (panel (d)) of the resonant conductance curve. The bright arcs visualized in all panels represent poles of the Green's function of the QD.}
\end{center}
\end{figure}

To shed more light on the effect of the tuning of charge and heat transport in the system, we make a plot of the quantity $\mathcal{T}=-\Gamma\text{{Im}}(\tilde{\mathcal{G}}_{d_{1}d_{1}})$ appearing in the Eq.\,(\ref{eq:G}) and Eq.\,(\ref{eq:Transmittance}), as function of $\varepsilon_{1}$ and $\varepsilon$, see Figs.\ref{fig:4}(a)-(d). Fig.\ref{fig:4}(a) corresponds to the case $t=\Delta,\varepsilon_2=0$. One can recognize a ``cat eye''-shaped central structure, corresponding to the vertical line at $\varepsilon=0$. Everywhere along this line $\mathcal{T}=\text{constant}$, which according to the Eq.~(\ref{eq:GK}) means that changes in $\varepsilon_1$ do not affect the conductance. This corresponds well to the conductance plateau in the Fig.~\ref{fig:2}. If $\varepsilon_2$ is finite, the ``cat eye'' structure transforms into a double-fork profile as it is shown in the Fig.\,\ref{fig:4}(b). Note that in this case, movement along the vertical line corresponding to $\varepsilon=0$ lead to the change of the function $\mathcal{T}$, which according to the  Eq.~(\ref{eq:GK}) leads to the modulation of the conductance. The maximal value is achieved at the point $\varepsilon_1=0$, which corresponds well to the resonant character of the curves shown in the Fig.\ref{fig:2}.  The introduction of the finite value of $\varepsilon_2$ and the asymmetry of the coupling between the QD and MFs ($t\neq\Delta$) leads to the shifts of the double-fork structure either upwards by $\varepsilon_1$ scale for $t>\Delta$ (panel (c), blueshift of the resonant curves in the Fig.\ref{fig:3}) or downwards by $\varepsilon_1$ scale for $t<\Delta$ (panel (d),  redshift of the resonant curves in the Fig.\ref{fig:3}). {It should be noted that similar results to the transmittance were reported both theoretically (Ref.[30]) and experimentally (Ref.[31]) for the geometry of a linear Kitaev wire with a QD attached to one of its ends placed between source and drain metallic leads. Differently from the case considered in our work, the authors account for the spin degree of freedom and particularly for Ref.[31], they evaluate the dependence of the conductance on the energy level of the QD and magnetic field, while we further analyze $\epsilon$ and asymmetry of couplings dependencies relevant for the understanding of the tuner regime. Despite the distinct geometry and spinless regime, our results and those reported in Refs.[30,31] are in good correspondence with each other, thus validating the mechanism pointed out in Refs.[30,32] of field-assisted overlapping between MFs and tunnel-couplings with the QD.}

\begin{figure}[!]
\begin{center}
\includegraphics[width=0.55\textwidth,height=0.3\textheight]{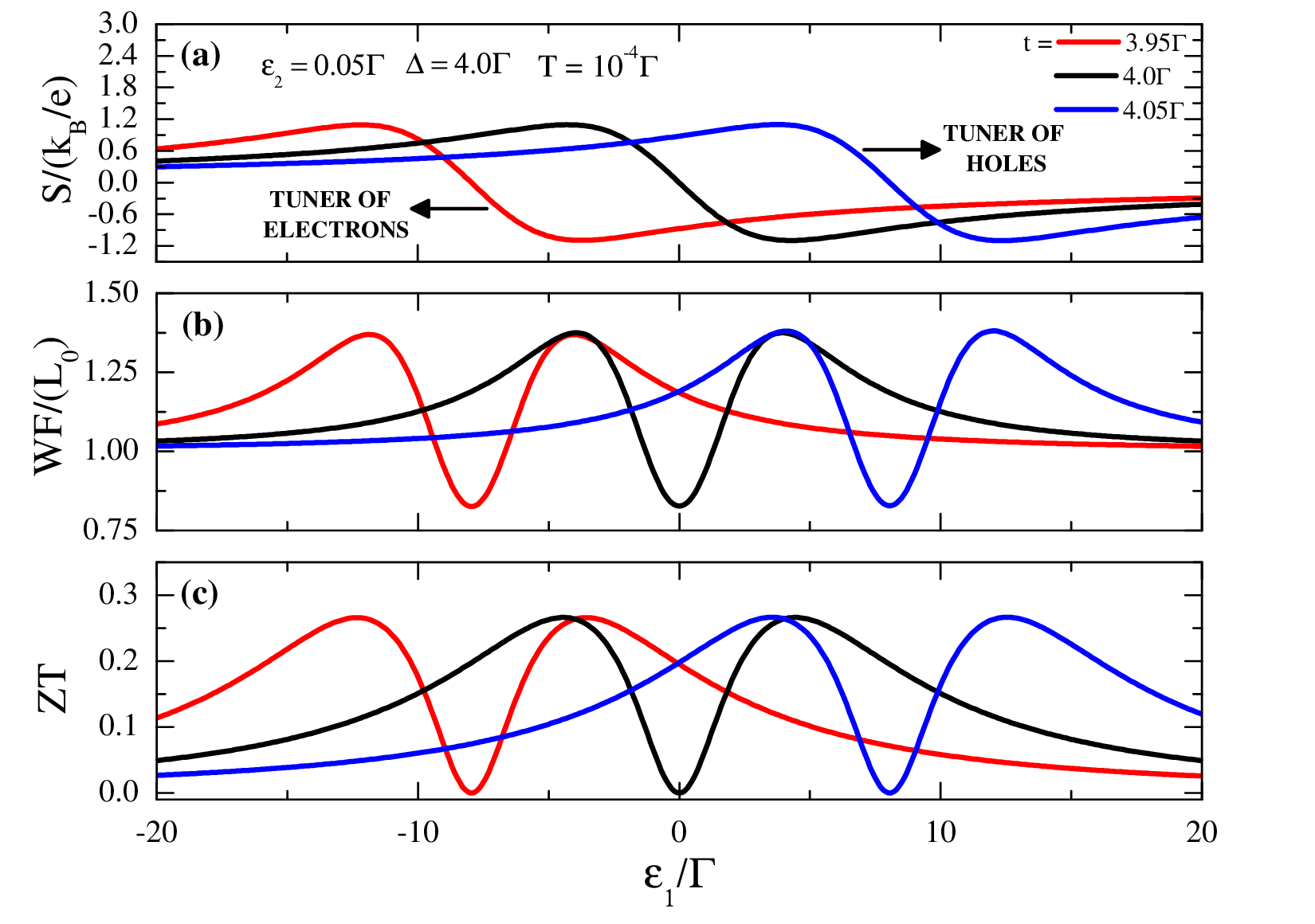}
\caption{\label{fig:5}(a) Thermopower $(S),$ (b) Wiedemann-Franz law $(WF)$ and (c) the figure
of merit $(ZT)$ as function of the QD energy level $\varepsilon_{1}$ for several $\varepsilon_{2}$
values of the couplings between MFs. Deviation from the condition $t=\Delta$ leads to the shift of the curves.}
\end{center}
\end{figure}

The possibility to tune electric and thermal conductances opens a way for tuning the thermopower $(S),$ Wiedemann-Franz law $(WF)$ and dimensionless figure of merit $(ZT)$ as it is shown in the Figs.\ref{fig:5}(a)-(c). In the Fig.\ref{fig:5}(a) the dependence of the thermopower on $\varepsilon_1$ is demonstrated. If $t>\Delta$, at $\varepsilon_1=0$, $S>0$ and the setup behaves as a tuner of holes. On the contrary, for $t<\Delta$, at $\varepsilon_1=0$, $S<0$ and the setup behaves as a tuner of electrons. Figs.\ref{fig:5}(b) and (c) illustrate the violation of $WF$ law and the behavior of the dimensionless thermoelectric $ZT$, respectively. Note that  ZT does not reach pronounced amplitudes, i.e, $ZT<1$\cite{Ventra}, even for finite values of $G$ and $K$ as dependence on $S^{2}$ prevails if we take into account Eq.\,(\ref{eq:GK}) into Eq.\,(\ref{eq:ZT}).

\section*{Conclusions}\label{sec4}

\hspace{0.3cm}In summary, we considered theoretically thermoelectric conductances for the device consisting of an individual QD coupled to both pair of MFs and metallic leads. The charge and heat conductances of this system as functions of an electron energy in the QD reveal resonant character. The position of the resonance can be tuned by changing the degree of asymmetry between the QD and the MFs, which allows us to propose the scheme of the tuner of heat and charge. Thermopower, Wiedemann-Franz law and the figure of merit are found to be sensitive to the asymmetry of the coupling as well. Our findings will pave way for the development of thermoelectric nanodevices based on MFs.

\section*{Acknowledgements}

This work was supported by the Brazilian funding agencies CNPq Grant No. 307573/2015-0, CAPES and S\~{a}o Paulo Research Foundation (FAPESP) Grant No. 2015/23539-8. I.A.S. acknowledges support from Horizon2020 RISE project CoExAN and the project No. 3.8884.2017/8.9 of the Ministry of Education and Science of the Russian Federation.

\section*{Author contributions}

A.C.S., M.S.F. and I.A.S formulated the problem and wrote the manuscript. L.S.R and A.C.S. derived the expressions and M.S.F. performed their numerical computing. F.A.D. and L.S.R. plotted the figures. All co-authors taken part in the discussions and reviewed the manuscript as well.

\vspace{0.3cm}

\noindent\textbf{Competing financial interests:} The authors declare no competing financial interests.


\end{document}